\documentclass{jors}

\definecolor{mygray}{gray}{0.6}

\usepackage[style=numeric,backend=biber,url=false,giveninits=true]{biblatex}
\usepackage{soul}
\DeclareNameAlias{default}{family-given}
\renewbibmacro{in:}{%
  \ifentrytype{article}{}{\printtext{\bibstring{in}\intitlepunct}}}

\begin{document}

{\bf Software paper for submission to the Journal of Open Research Software} \\



\rule{\textwidth}{1pt}

\section*{(1) Overview}

\vspace{0.5cm}

\section*{Title}
Epidemiological Agent-Based Modelling Software (Epiabm)

\section*{Paper Authors}
1. Gallagher, Kit;\textsuperscript{\dag{}} \\
2. Bouros, Ioana;\textsuperscript{\dag{}} \\
3. Fan, Nicholas;\textsuperscript{\dag{}} \\
4. Hayman, Elizabeth;\textsuperscript{\dag{}} \\
5. Heirene, Luke;\textsuperscript{\dag{}} \\
6. Lamirande, Patricia;\textsuperscript{\dag{}} \\
7. Lemenuel-Diot, Annabelle; \\
8. Lambert, Ben; \\
9. Gavaghan, David; \\
10. Creswell, Richard\\ \\
\textsuperscript{\dag{}}These authors contributed equally to this research.

\section*{Paper Author Roles and Affiliations}
1. Mathematical Institute, University of Oxford, UK \\
2. Department of Computer Science, University of Oxford, UK \\
3. Mathematical Institute, University of Oxford, UK \\
4. Department of Engineering Science, University of Oxford, UK \\
5. Mathematical Institute, University of Oxford, UK \\
6. Mathematical Institute, University of Oxford, UK \\
7. Roche Pharmaceutical Research and Early Development, Pharmaceutical Sciences, Roche Innovation Center Basel, Switzerland \\
8. College of Engineering, Mathematics and Physical Sciences, University of Exeter, UK \\
9. Department of Computer Science, University of Oxford, UK \\
10. Department of Computer Science, University of Oxford, UK

\section*{Abstract}

Epiabm is a fully tested, open-source software package for epidemiological agent-based modelling, re-implementing the well-known CovidSim model from the MRC Centre for Global Infectious Disease Analysis at Imperial College London. It has been developed as part of the first-year training programme in the EPSRC SABS:R$^3$ Centre for Doctoral Training at the University of Oxford. The model builds an age-stratified, spatially heterogeneous population and offers a modular approach to configure and run epidemic scenarios, allowing for a broad scope of investigative and comparative studies. Two simulation backends are provided: a pedagogical Python backend (with full functionality) and a high performance C++ backend for use with larger population simulations. Both are highly modular, with comprehensive testing and documentation for ease of understanding and extensibility. Epiabm is publicly available through GitHub at \href{https://github.com/SABS-R3-Epidemiology/epiabm}{\url{github.com/SABS-R3-Epidemiology/epiabm}. }

\section*{Keywords}
Agent-based modelling; epidemiology; COVID-19; Python; C++

\section*{Introduction}
As a result of the recent COVID-19 pandemic, a wealth of epidemiological models have been developed, or adapted from pre-existing models. These range from simple, population-wide models which predict the total number of infected individuals in a population \cite{He2020}, to models with spatial variation/age stratification to track the case rates among different groups within a population \cite{Wu2011, Lyra2020}. Some of the most complex models, often used to advise government and policy-maker decisions, are agent-based models (ABMs) \cite{Hunter2018}. The representation of populations on an individual scale by ABMs allows for the modelling of complex spatial and behavioural phenomena which are not straightforward to account for in traditional population-averaged, compartmental differential equations models. In these spatially homogeneous models, only the total number of individuals at each step of disease progression (e.g. Susceptible, Infected, Recovered) is modelled. In contrast, implementing individual agents allows for complex inter-agent interaction networks with more realistic transmission dynamics, where interaction probabilities can be affected by each individual's spatial position and age.\\

One of the most influential models in the UK during the COVID-19 pandemic has been the CovidSim model \cite{Ferguson2020_CovidSim}, developed by the MRC Centre for Global Infectious Disease Analysis at Imperial College London. The model was initially designed to support influenza pandemic planning \cite{Ferguson2006}. At the start of the COVID-19 pandemic, the code was very rapidly adapted to enable modelling of the initial stages of the outbreak. Notably, the model was used to produce the high profile ``Report 9'', which considered the impact of various non-pharmaceutical interventions (NPIs) on the transmission of COVID-19 \cite{Ferguson2020_Report9}, and this report is widely held to have been influential in the UK government's decision making \cite{Adam2020}. This model has a higher level of spatial and behavioural complexity than alternative age-stratified models \cite{Birrell2021_PHEmodel, Danon2021_MetaWards, Keeling2021_WarwickModel, Davies2020_CMMID}.\\

CovidSim is highly efficient allowing large-scale simulations to be run. The model and code were originally created to underpin a time-critical academic publication. Because of this, the codebase is not (and was not intended to be) easily extensible. A range of changes to the code and software engineering practices have therefore been suggested in the academic literature, including making it more modularised; controlling the random number seeding to make results entirely reproducible; and including comprehensive code testing and documentation \cite{Shen2020,Wooldridge2020_blog, Eglen2020_Codecheck}.\\

The development of a fully modular and well-documented version of CovidSim code would therefore have multiple potential benefits both for modelling and responding to future pandemics, and as a pedagogical tool. The modularity will enable researchers across the epidemiological modelling community to isolate and characterise dominant transmission mechanisms and viral characteristics in agent-based models, and allow flexible configuration of interventions. In this paper, we present such a modular and fully documented re-implementation of the CovidSim model, which we call \textit{Epiabm}, that adheres to professional software development principles \cite{Crick2017, Stodden2014}, for use in research and education settings. The Epiabm code has been developed as part of the first year training programme for PhD students at the \emph{EPSRC CDT in Sustainable Approaches to Biomedical Science: Responsible and Reproducible Research (SABS:R${^3}$) CDT} at the University of Oxford. The SABS:R${^3}$ programme focuses on providing comprehensive training in software development and software engineering to all of its students, in the context of industrially derived research in the biomedical sciences. To make this training immediately relevant, all students undertake an industry-supported group software development project over the course of their first year, allowing them to learn and apply their software skills in a realistic setting; the Epiabm software is the result of one of these year-long projects. The project was supported by colleagues in Roche's Infectious Disease Modelling Group. \\ 

Two simulation backends are provided: a pedagogical Python backend (with full functionality) and a high performance C++ backend for use with larger population simulations. Both are highly modular, with comprehensive testing and documentation for ease of understanding and extensibility. Epiabm is publicly available through GitHub at \url{github.com/SABS-R3-Epidemiology/epiabm}. 

\section*{Implementation and architecture}

\begin{figure}
    \centering
    \includegraphics[trim={2.5cm 3cm 3cm 2cm}, clip, width = 1.0\linewidth]{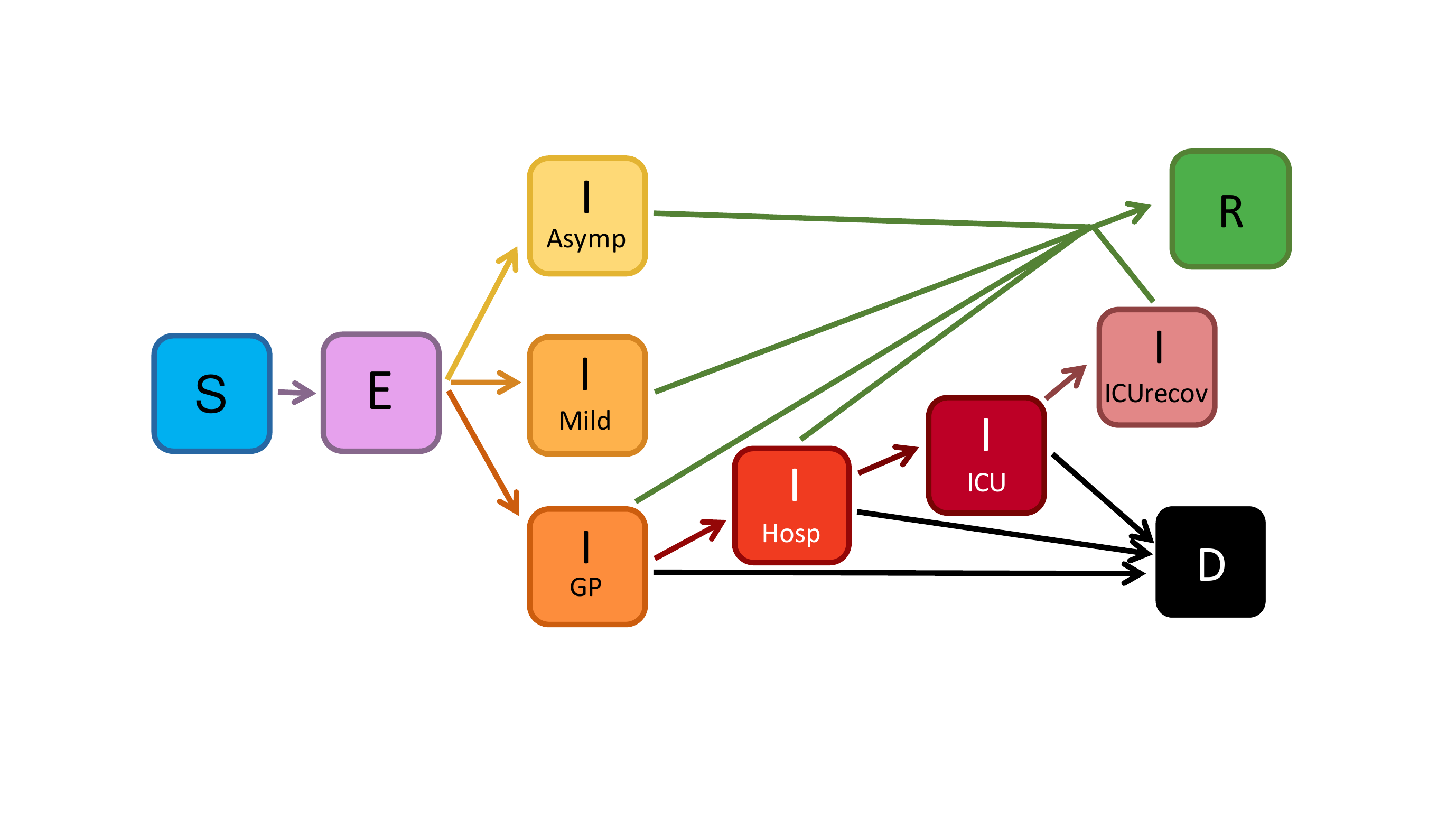}
    \caption{Infection state progression, with arrows depicting the possible routes of progression through this scheme. Based on a SEIRD model, where all individuals are either susceptible (S) to the disease, exposed (E) to the disease, infected (I) by the disease, or who have recovered (R) or died (D) from the disease. Multiple compartments for infection account for cases of differing severity, and this network is highly configurable, and new compartments or connections can be added with ease.}
    \label{fig:infection_network}
\end{figure}

The basic units of an agent-based model are individual people. For each individual, personal characteristics such as age are used to determine their vulnerability to infection. Each individual is also assigned a time-dependent infection state, which is based on the basic SEIR (Susceptible/Exposed/Infected/Recovered) model but can be extended to include additional infection states and sub-states (e.g. hospitalised individuals). Individuals may move between states according to a network defined by a state transition matrix, depicted in Figure \ref{fig:infection_network}. The set of states and the network between them used within the CovidSim model is implemented by default in Epiabm, but specific methods to add user-defined compartments and connections allow for configurability to allow modelling of other diseases or alternative networks of states.\\

All individuals are initially `Susceptible', a state where they have not been infected, and have no prior immunity to the virus. Upon being infected, individuals initially become `Exposed' (where they remain unable to infect others), before transitioning to an infected state after a randomly sampled latent time. The model implements a number of different `Infected' states with differing degrees of severity, where vulnerable individuals have a higher likelihood of entering states with increased probabilities of death. This method also allows pressure on public health systems and intensive care units to be tracked, through specific states for those ill enough to require these facilities. Individuals with sufficiently severe infections may subsequently die, while all others will enter the `Recovered' status after a randomly sampled waiting time. While this model does not include waning immunity by default, the configurability of this method allows for simple addition of a `Recovered' to `Susceptible' pathway over sufficient time. \\

\begin{figure}
    \centering
    \includegraphics[width = \linewidth]{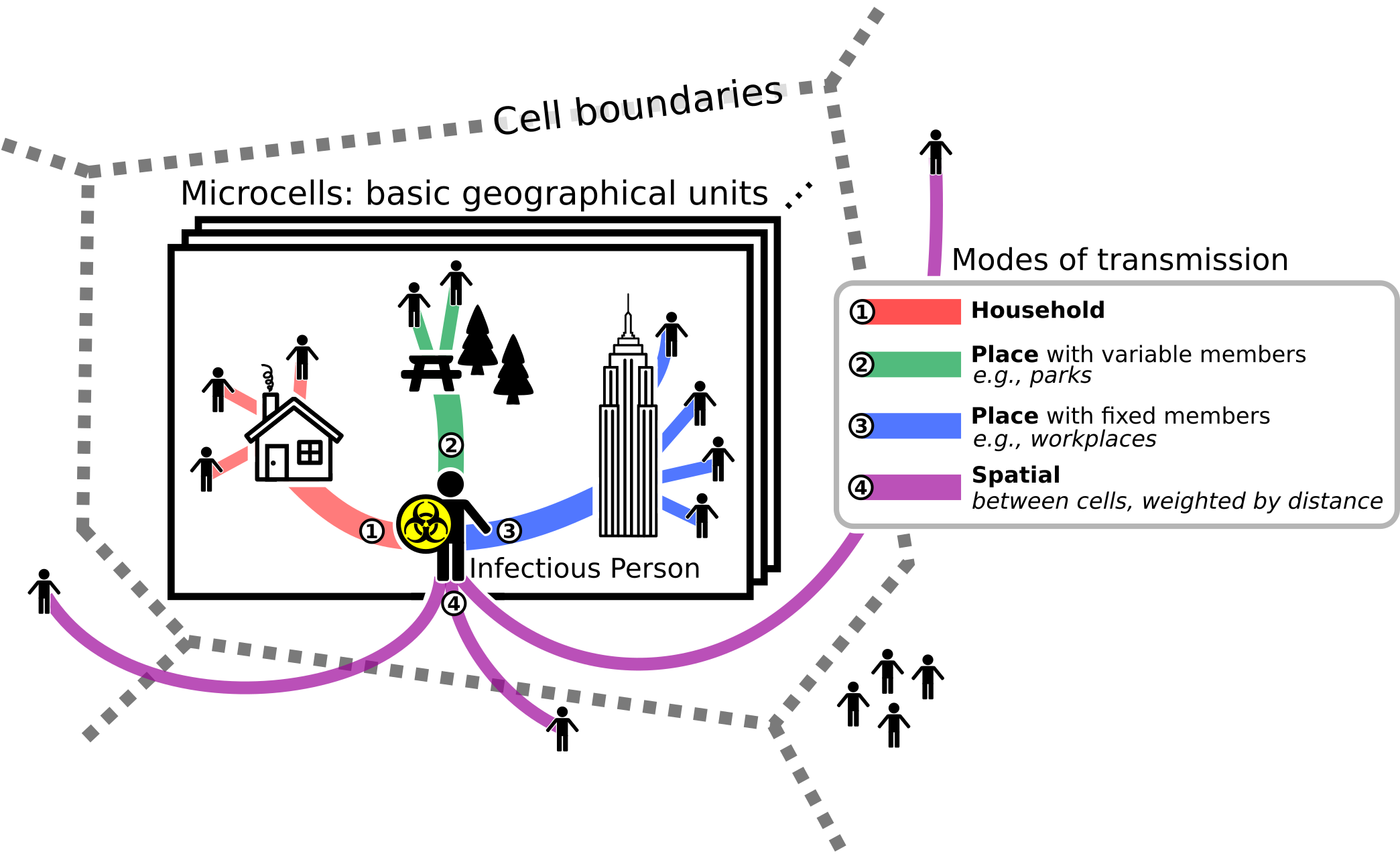}
    \caption{Population structure in Epiabm. A population is formed of many microcells, which are grouped into cells (the largest spatial unit). Infected individuals may infect others within their cell through households and places, while infections are spread between cells according to a spatial kernel.}
    \label{fig:population_structure}
\end{figure}

Each individual is a member of a ``population", with the overall population being made up of a hierarchy of cells and microcells as illustrated in Figure \ref{fig:population_structure}. Individuals are assigned to households and to places (such as schools or workplaces) within these microcells, through which they may infect other individuals. Such infection events are probabilistic and depend on a number of factors including the age and infectiousness of the individuals involved. \\

Epiabm supports a complete epidemic simulation through a number of key workflow steps:
\begin{itemize}
    \item Population Generation,
    \item Simulation Configuration,
    \item Simulation Evaluation,
    \item Results Output.
\end{itemize}

It also includes a range of example plotting scripts to visualise output data. Each of these workflow steps is explained below:\\

\textbf{Population Generation}. To generate a population, users can read in an external file with counts of places, households and individuals in each infection state per microcell (examples of this are provided in our repository). Alternatively, users may configure the population randomly based on given parameters (such as the number of places of each type per cell).\\

\textbf{Simulation Configuration}. A simulation may be configured by assigning a number of sweeps. These are functions that iterate over the population and are responsible for within-host infection progression and infection events between individuals via various transmission mechanisms (e.g.\ via households or places). A parameters file must also be specified here, with key-value pairs for each parameter used in the model - the default values provided are the same as those used in CovidSim, and referenced in our \href{https://github.com/SABS-R3-Epidemiology/epiabm/wiki}{Wiki}. \\

\textbf{Simulation Evaluation}. To evaluate the simulation, the sweep functions are called at each time step. This is completely modular, and so different infection mechanisms (such as via households or workplaces) may be removed independently to explore the role of different transmission mechanisms in the viral spread. \\

\textbf{Results Output}. At each timestep, output loggers may be used to record the current state of the simulation in a .csv file, over a range of resolutions -- from microcell to global.\\ 

A full representation of the simulation routine is given in Appendix \ref{app:Epiabm_pseudocode}.

\subsubsection*{Comparison to CovidSim}

While we have endeavoured to emulate both the overall structure and the functionality of CovidSim, the architecture we have used to achieve this in Epiabm differs significantly. Most notably, we have used a strongly object-oriented approach to population generation and storage; while less efficient, we believe this is more intuitive and will enable other users to easily adapt sections of the code for their own use. While many aspects of simulation configuration (such as determining which infection sweeps to use) are specified through command line flags and parameter files in CovidSim, we have chosen to manage configuration though workflow scripts, again to increase readability and ease of sharing. This is made possible through our modular architecture; while CovidSim combines sweeps and output functions in large code blocks, Epiabm has separate individual classes and methods that may be included or excluded at will. Related classes, such as the different infection sweeps, also inherit from abstract parent classes to reduce code duplication where daughter classes have similar functionality.

\subsubsection*{Example Simulation}
\begin{figure}[ht]
    \centering
    \includegraphics[width = 0.8\linewidth]{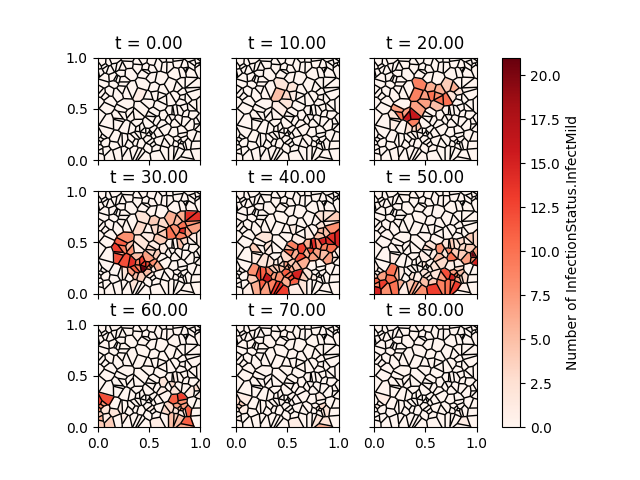}
    \caption{Spatial Distribution of infected individuals within the population, at different time points during the simulation. Configured with a population of 10,000 people distributed across 200 cells, each containing 2 microcells with 5 households per microcell. One infected individual is initialised in the central cell, with mild infection status; the simulation is run for 80 days. Inter-cell infections only occur between nearby cells, allowing visualisation of the infection propagating through the simulation region over time.}
    \label{fig:example_simulation}
\end{figure}

An example simulation was configured using a synthetic population of 10,000 individuals distributed across 200 cells, each containing 2 microcells with five households per microcell. One infected individual is initialised in the central cell, with mild infection status. The resulting propagation of the infection through the population over a period of 80 days is visualised in Figure \ref{fig:example_simulation}.\\ 

\begin{figure}
\centering
\begin{subfigure}[t]{.5\textwidth}
    \centering
    \includegraphics[width = \linewidth]{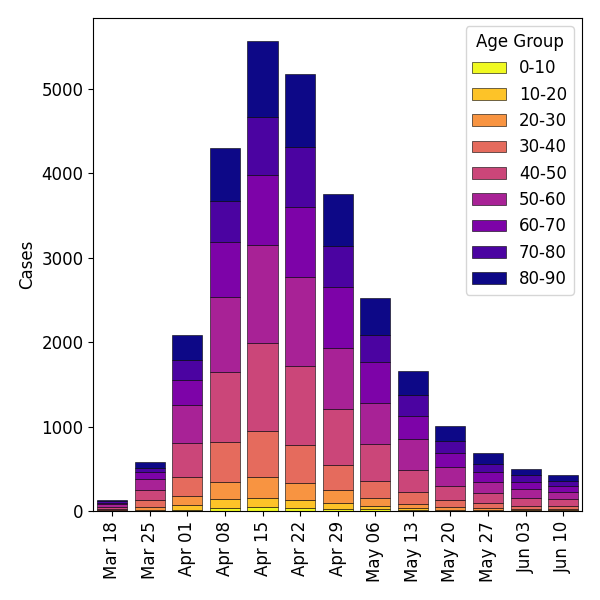}
    \caption{pyEpiabm}
    \label{fig:Gibraltar_Epiabm}
\end{subfigure}%
\begin{subfigure}[t]{.5\textwidth}
  \centering
  \includegraphics[width=\linewidth]{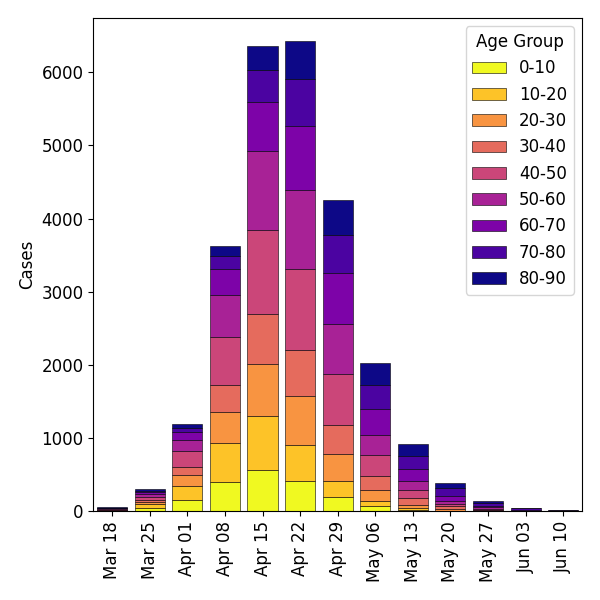}
  \caption{CovidSim}
  \label{fig:Gibraltar_CovidSim}
\end{subfigure}
\caption{A comparison of simulation outputs from pyEpiabm \textbf{(a)} and CovidSim \textbf{(b)}, for an epidemic in Gibraltar initiated by 100 infected individuals. While the outputs are highly stochastic, a strong agreement is broadly observed.}
\label{fig:Gibraltar_Comparison}
\end{figure}

We also configured a national-scale simulation based on real-world parameter values to enable direct comparison to the results of CovidSim. The region of Gibraltar (with an approximate population of 34,000) was chosen for computational simplicity. Age-stratified output plots from both software packages are displayed in Figure \ref{fig:Gibraltar_Comparison} and show strong agreement, with the number of weekly cases peaking around April 15--April 22.\\

This simulation takes 42 seconds to run on the python backend, and 8 seconds on the C++ backend (including all population configuration and model running) on an AMD 3600X processor (6 cores, 3.8 GHz). In comparison, CovidSim takes a total of 45 seconds, although this is heavily dominated by build time (which does not scale as heavily in more complex cases) with the simulation alone running in 4.5 seconds. This highlights the performance compromises that have been necessary particularly in pyEpiabm, to ensure modularity and readability, relative to the highly optimised codebase of CovidSim. Despite this, we are able to perform complete epidemic simulations on small countries on a personal computer in feasible time periods, sufficient for many educational and research purposes.

Full code listings for these simulations are available on \href{https://github.com/SABS-R3-Epidemiology/epiabm/tree/main/python_examples}{GitHub}, alongside more basic workflows to introduce new users to the simulation capabilities of this software.

\section*{Quality control}

Both the Python and C++ backends have full unit testing with 100\% test coverage, verifying expected behaviour of all deterministic and stochastic methods. Functional testing is used for automatic verification of random seed reproducibility in both population generation and simulation methods. \\

Epiabm also has testing routines to ensure all publicly exposed methods and classes are included in the documentation, and uses Flake8 linter tests to ensure that contributed code is consistent in style. All these tests are included in a continuous integration pipeline implemented through GitHub workflows, with unit testing evaluated across Python versions 3.6-3.9 and the latest macOS, Windows and Ubuntu distributions with Python version 3.8.\\

The CONTRIBUTING.md file in the epiabm repository contains more detailed and up-to-date information on our development workflow, testing and CI infrastructure, and coding style guidelines.\\

Users are also provided with a number of example workflows, to configure and run different types of simulation, as well as plotting methods to visualise their outputs. This includes simple simulations with known behaviour, to allow bench-marking against pre-existing models.\\

\section*{(2) Availability}
\vspace{0.5cm}
\section*{Operating system}

Epiabm uses no functions specific to any operating system (OS) and so can run on any OS that provides Python and C++.

\section*{Programming language}

\begin{itemize}
    \item Python -- version 3.6 or higher.
    \item C++ -- version 17 or higher (for cEpiabm only)
    \item CMake -- version 3.15 (for cEpiabm only)
\end{itemize}

\section*{Additional system requirements}

Memory and disk space dependent on usage case.

\section*{Dependencies}

Essential:
\begin{itemize}
    \item numpy $>=1.8$
    \item packaging
    \item pandas $>=1.4$
    \item tqdm
\end{itemize}

Optional:
\begin{itemize}
    \item flake8 $>= 3$ -- Used to check code style
    \item matplotlib -- Used in example workflow
    \item parameterized -- Used in unit tests
    \item sphinx $>= 1.5$ -- Used to generate documentation
\end{itemize}

\section*{List of contributors}
Our project is hosted on GitHub, and publicly visible, allowing researchers from around the world to find our work and contribute to the codebase.

\begin{itemize}
    \item Open-Source Software Contribution -- Saket Kumar, Netaji Subhash Engineering College, Maulana Abul Kalam Azad University of Technology, India.
    \item Open-Source Software Contribution -- Pietro Monticone, University of Turin.
\end{itemize}

\section*{Software location:}

{\bf Archive}

\begin{description}[noitemsep,topsep=0pt]
	\item[Name:] Zenodo
	\item[Persistent identifier:] DOI: 10.5281/zenodo.7327444
	\item[Licence:] BSD 3-Clause
	\item[Publisher:]  Kit Gallagher
	\item[Version published:] 1.0.1
	\item[Date published:] 16/11/22
\end{description}

{\bf Code repository}

\begin{description}[noitemsep,topsep=0pt]
	\item[Name:] GitHub
	\item[Persistent identifier:] \url{https://github.com/SABS-R3-Epidemiology/epiabm}
	\item[Licence:] BSD-3-Clause
	\item[Date published:] 01/03/22
\end{description}

\section*{Language}

English\\
\\

\section*{(3) Reuse potential}

Epiabm is designed with both research and educational use in mind. The modular aspect of the code allows for highly configurable simulations and investigation into the sensitivity of large scale agent-based models to different transmission mechanisms. While the default parameter values provided are tailored to the spread of COVID-19 within the UK population, minimal reconfiguration is required to adapt these for other countries/diseases. The modular nature of the code offers considerable freedom to explore and compare the roles of different transmission mechanisms in epidemic growth for viral strains with varied properties. The model can also be extended to include custom interventions, both pharmaceutical and non-pharmaceutical, on local and global spatial scales, and indeed this will be the task of one the first year group projects in the coming academic year.\\

Detailed documentation and example workflows are provided, also enabling use of Epiabm in educational settings and for users with little familiarity with agent-based epidemiological models. We welcome questions, suggestions, bug reports, and user contributions via the GitHub repository, which acts as a central communication platform for Epiabm. A detailed guide on contributing to Epiabm is also available there. 

\section*{Acknowledgements}

We acknowledge the advice of Fergus Cooper in setting up the C++ software architecture.
We also acknowledge the work of Martin Robinson from the University of Oxford, as well as Steve Crouch and James Graham from the Software Sustainability Institute, for their helpful instruction on the topic of software engineering.

\section*{Funding statement}

All authors except A.L.D. acknowledge funding from the EPSRC CDT in Sustainable Approaches to Biomedical Science: Responsible and Reproducible Research - SABS:R3 (EP/S024093/1). R.C. acknowledges funding from a doctoral training partnership studentship in the Department of Computer Science at the University of Oxford. A.L.D. was funded as a Roche Pharmaceutical Research employee.

\section*{Competing interests}

The authors declare that they have no competing interests.


\printbibliography

\begin{appendices}
\section{Epiabm Implementation} \label{app:Epiabm_pseudocode}
This section introduces our implementation of CovidSim, in the pyEpiabm backend. The cEpiabm backend broadly follows the same structure at all levels, and all pseudocode given here also applies to this backend.

The overall workflow is illustrated in Algorithm \ref{alg:epiabm_workflow}, which represents the simulation flow file and the broad functionality of the `\texttt{Simulation}' class. The bulk of the simulation work occurs within the `\texttt{sweep()}' function calls, an example of which is given in Algorithm \ref{alg:example_sweep}.

\begin{algorithm}
\caption{Epiabm Workflow}\label{alg:epiabm_workflow}
\begin{algorithmic}
\Require parameters.json  file
\Ensure Output .csv files   \Comment{Containing case numbers at each timestep}
\State pop\_params $\gets$ \{`population\_size': $1000$, ...\}   
\State population $\gets$ ToyPopulationFactory.make\_pop(pop\_params)
\State{}
\State sim $\gets$ Simulation()
\State sim.configure(population, list\_of\_sweeps)

\State t $\gets$ start\_time
\While{t $<$ end\_time}
    \State HouseholdSweep()
    \State PlaceSweep()
    \State t $\gets$ t $+$ time\_step
\EndWhile

\end{algorithmic}
\end{algorithm}

\begin{algorithm}
\caption{Example epiabm sweep - HouseholdSweep()}\label{alg:example_sweep}
\begin{algorithmic}
\Require population, t (current time)
\Ensure Queues individuals newly infected within households
\For{cell in population.cells}
    \For{infector in cell.persons}
        \State \textbf{assert} infector.household is not None:
        \If{not infector.is\_infectious()}
            \State \textbf{continue}
        \EndIf
        \State
        \For{infectee in infector.household.persons}
            \If{not infectee.is\_susceptible()}
                \State \textbf{continue}
            \EndIf

            \State force\_of\_infection $\gets$ household\_foi(infector, infectee, t)

            \State r $\gets$ random.uniform($0, 1$)
            \If{r $<$ force\_of\_infection}
                \State cell.enqueue\_person(infectee)   \Comment{Queue person for infection}
            \EndIf
        \EndFor
    \EndFor
\EndFor
\end{algorithmic}
\end{algorithm}
\end{appendices}

\vspace{2cm}

\rule{\textwidth}{1pt}

{ \bf Copyright Notice} \\
Authors who publish with this journal agree to the following terms: \\

Authors retain copyright and grant the journal right of first publication with the work simultaneously licensed under a  \href{http://creativecommons.org/licenses/by/3.0/}{Creative Commons Attribution License} that allows others to share the work with an acknowledgement of the work's authorship and initial publication in this journal. \\

Authors are able to enter into separate, additional contractual arrangements for the non-exclusive distribution of the journal's published version of the work (e.g., post it to an institutional repository or publish it in a book), with an acknowledgement of its initial publication in this journal. \\

By submitting this paper you agree to the terms of this Copyright Notice, which will apply to this submission if and when it is published by this journal.

\end{document}